\documentclass[%
 reprint,
 amsmath,amssymb,
 aps,
]{revtex4-2}

\usepackage{graphicx}
\usepackage{dcolumn}
\usepackage{bm}
\usepackage{float}
\usepackage{comment}
\usepackage{subfigure}
\usepackage{multirow}
\usepackage{setspace}
\usepackage{lineno}

\begin{document}

\preprint{APS/123-QED}

\title{Dark Matter Induced Neutron Production Search Limits}

\author{Haichuan Cao}%
\author{David Koltick}
 \altaffiliation[Corresponding Author: koltick@purdue.edu]{}
\affiliation{Department of Physics and Astronomy\\ 525 Northwestern Ave., West Lafayette, Indiana USA, 47907}





\date{\today}

\begin{abstract}
\noindent
An independent indirect detection search for Dark Matter-Matter (DM-M) interactions is undertaken to set cross section limits based on neutron production data collected  by the NMDS-II detector for  1440 hours at 1166 m.w.e. and 6504 hours at  583 m.w.e.. The detector system consists of a 30 cm cube Pb-target instrumented with 60 $^3$He neutron counters. The neutron detector system calibrated with a $^{252}$Cf source  yields a single particle detection efficiency of 23.2\%$\pm$1.2\%.    During data collection, the highest neutron multiplicity event observed 54 neutrons.  The neutron multiplicity, n, distribution,  fits well to a power law $k \times n^{-p}$, for both the data and cosmic ray muon induced neutron production in Geant4 simulations. Two DM-M interaction models were used to set limits. The first, a spallation model, assumes a single proton with kinetic energy equal to the DM-M interaction energy. The other, a fire-ball model assumes an annihilation between DM-M producing pions with a limiting Hagedorn temperature. The two extreme models produce similar upper DM-M cross section limits over the DM mass range between 300 MeV to 100 GeV.  Limits assume all the DM energy is deposited in the Pb-target. Spin independent limits, proportional to A$^{-2}$, are at the level ~10$^{-45}~ cm^2$.  Spin dependent limits, proportional to A$^{-1}$, are at the level, $2\times 10^{-42}~cm^2$.
\end{abstract}

\maketitle


\section{Introduction}

The nature of dark matter has been one of the most intriguing problems in fundamental physics and cosmology. Based on the standard model of cosmology, the total mass-energy of the known universe contains 4.9\% ordinary matter, 26.8\% dark matter and 68.3\% dark energy \cite{particle2022review}. Although there is extensive evidence of dark matter in astrophysical observations, dark matter particles have not been detected. Dark matter searches can use direct or indirect detection methods. 

 Present WIMP Dark Matter (DM) search strategies are focused on possible direct detection through elastic or inelastic scatterings on atomic nuclei, or with electrons. This approach become insensitive to M$_{DM} <~$10 GeV. The XENON series \cite{aalbers2022next}, LUX \cite{akerib2021effective} and SuperCDMS \cite{albakry2022strategy} are some notable dark matter direct detection experiments. For mass $>10$ GeV direct detection methods \cite{akerib2017results, cui2017dark, collaboration2018dark} yield  DM-nucleon cross section limits of order $\sim 10^{-46}~cm^2$.

Indirect DM detection refers to the search for the annihilation or decay debris from DM particles, resulting in detectable species, including especially gamma rays, neutrinos, and antimatter particles. Astrophysical observations play a dominant role in current Dark Matter
indirect detection searches. Examples are gamma observations from the Cherenkov Telescope Array\cite{doro2013dark}, neutrino observation from  IceCube \cite{abbasi2023search}, and results from the ANTARES Telescope \cite{adrian2016limits}.  

 However, for low-mass dark matter particles, $m < 10\ GeV$, the sensitivities for DM-M or DM-DM couplings are poor in both current direct detection searches and indirect detection searches. This situation can be improved by the search for DM-M interactions observable through excess high multiplicity neutron production in nuclear targets.  Even for M$_{DM} <$ 10 GeV DM-M annihilation is capable of producing large signals, $>$200 neutrons if the energy is deposited in a Pb-target. For high energy particles, a solid detector consisting of high mass number nuclei, acts similar to a solid state detector to $\gamma$-rays, except the band gap is of order $\sim$10 MeV. In addition, DM-M spin independent cross sections are proportional to $A^2$ and A for spin dependent interactions, favoring a   simple Pb-target. 
 
 For these reasons we reanalyzed the Neutron Multiplicity Detector System-II (NMDS-II) data sets (2001- 2003) for an indirect dark matter search using a Pb-target. A unique feature of the data sets is the identical detector being located underground at both 583 m.w.e. and 1166 m.w.e.. Such a placement allows a reduction in the systematic errors by comparing the results to one another. In addition, the higher level of cosmic ray induced background at the 583 m.w.e. depth is used to verify the background expectations at 1166 m.w.e. depth. It is the data set at 1166 m.w.e. depth that is used to set the DM interaction cross section limits.

Limits are set based on the background neutron number spectrum comparing the two  NMDS-II  data sets with Geant4 simulations of cosmic ray muon induced neutron production, both directly by muons passing through the detector as well as by interactions in the detector from associated muon induced shower particles produced in the surrounding rock.

\section{The NMDS-II Experiment}

\subsection{Neutron Detection System}

\begin{figure}
	\centering
	\includegraphics[width=0.75\linewidth]{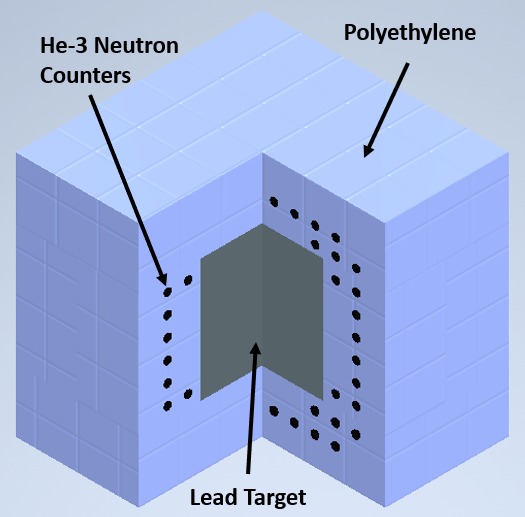}
	\caption{The NMDS-II detector, consists of a 30 cm cube Pb-target surrounded by 60 $^3$He detectors encased in polyethylene. The 60 cm polyethylene cube is composed of 4 side panels and two cap panels. The distribution of $^3$He tubes is displayed using a cross-section drawing. }
	\label{fig:CUPPdetector}
\end{figure}

The NMDS-II detector is an instrumented 30 cm cube of Pb (305 kg), which serves as the target. The Pb-cube is surrounded by 15 cm thick polyethylene moderators, within which 60 $^3$He proportional neutron counters, model SNM-18, produced by Maxiums Energy, are encased on all 6-sides, as shown in the Figure \ref{fig:CUPPdetector}. The $^3$He proportional tubes serve as the neutron multiplicity detection system. 

\begin{figure}
	\centering
	\includegraphics[width=0.8\linewidth]{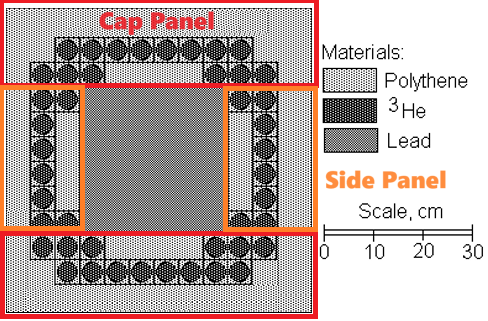}
	\caption{Cross section through the middle of the NMDS-II  detector showing the $^3$He neutron counters. The dimensions of the $^3$He tubes are not in proportion. The top and bottom cap panels are outlined in red.  The side panels are outlined in orange. }
	\label{fig:arrange_He3}
\end{figure}

 Each counter measures 28.5 cm long and 1.55 cm in diameter filled with a mixture of $^3$He (75\%) and Ar(25\%) at 4 atm pressure and operated in proportional mode at 1400 V. The neutron counters are arranged in top and bottom cap panels and 4 side panels forming the polyethylene box as shown in the Figure \ref{fig:arrange_He3}. The positions of the $^3$He counters were chosen by Monte-Carlo modeling, optimized so that neutron events occurring at any point in the target would be registered with almost the same efficiency.

A single neutron hit triggers the $^3$He counter system to record data for 256 $\mu s$. Each hit causes the tube to be non-responsive or dead for a period of 10 $\mu s$. The neutron half-life in the lead target is 65 $\mu s$, resulting in an average of 6.47\% of the total neutrons produced to be outside the data collection window.

\begin{figure}

\centering
\includegraphics[width=0.9\linewidth]{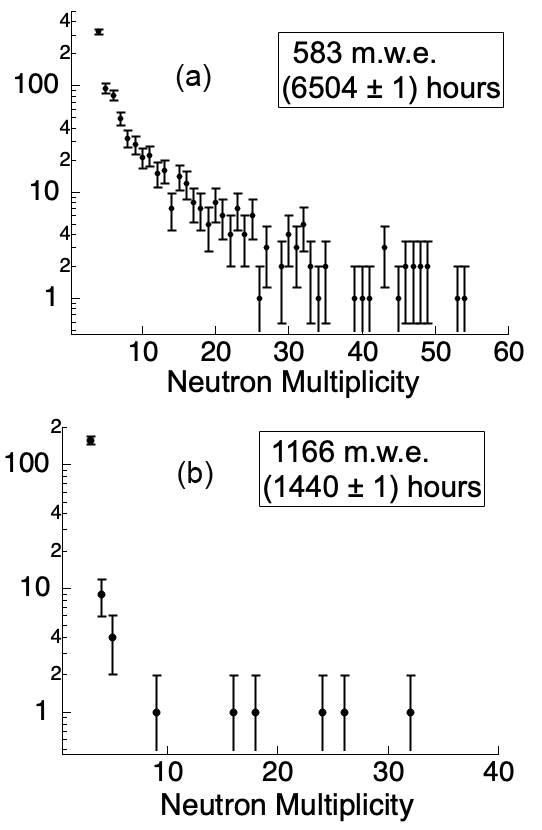}

\caption{Observed neutron multiplicity distributions at (a) 583 m.w.e.  and at (b) 1166 m.w.e. .}
\label{fig:CUPPdata}
\end{figure}

The two neutron multiplicity data sets collected by the NMDS-II, at 583 m.w.e. and at 1166 m.w.e. are shown in Figure \ref{fig:CUPPdata}.

\subsection{Environmental Background Rejection}

Apart from cosmic-induced neutrons, the main background in the NMDS-II experiment is radioactivity inside the lead target and from the surrounding rock. Neutrons arise from two sources: i) spontaneous fission of $^{238}$U and ii) ($\alpha, n$) reactions initiated by $\alpha$-particles from U or Th decay \cite{Kudryavtsev:2008fi}.  

Independent Geant4 simulations for the two processes  were performed to estimate the upper limit of the number of neutrons detected due to natural neutron radioactivity. The $n \rightarrow 2n$ process with threshold at $E_n \sim 7.5 MeV$ and the $n \rightarrow 3n$ process with threshold at $E_n \sim 16 MeV$ in the lead target were included. For the spontaneous fission, the neutron energy spectrum is obtained from \cite{lovchikova2004spectra} and the number distribution of fission neutron is obtained from \cite{lestone2014comparison}. For ($\alpha, n$) reactions, the neutron energy is set to be uniformly distributed from 0 to 12 MeV, taken from the upper limit of the neutron energy generated from the Be($\alpha, n$)C reaction. Doing so overestimates the mean neutron energy, and likewise the $n \rightarrow 2n$ rate is overestimated.  

 At 1166 m.w.e., 157 events with 3 detected neutrons are observed during 1440 hours. These includes electronic noise events, natural decays, and cosmic-ray-produced neutrons. This observations can be used to set an upper limit on the contribution of natural decay induced events populating the analysis sample. Assuming all events with detected neutron multiplicity equal to 3 originate from $^{238}U$ fission inside the lead target, the simulation yields 9 events at the detected multiplicity 4 and 0.2 events at detected multiplicity 5. Assuming all events with detected multiplicity 3 originate from ($\alpha, n$) inside the lead target, the simulation yields less than 0.01 events at the detected multiplicity at 4 and less than 0.001 events at the detected multiplicity at 5. 

 The same study was conducted for the production of $^{238}U$  fission and ($\alpha$, n) from surrounding rock, also matching 157 events with 3 detected neutrons.  The $^{238}U$ fission yielded less than 0.05 events with 5 detected neutrons or more and ($\alpha$, n) yielded less than 0.001 events with 5 detected neutrons or more.

In addition, there are system-wide single-rate backgrounds, including electronic noise. At 1166 m.w.e., the observation rate of at least 1 neutron is $\sim 0.6\ n/s$, which includes electronic noise, natural decays and cosmic ray produced neutrons. This rate can be used to set an upper limit on the single-rates producing false higher neutron multiplicity events. If it is assumed that the number of events in one time selection window, $256 \mu s$, follows a Poisson distribution, the probability of observed neutron multiplicity at 4 or more is less than $10^{-6}$. 

With these considerations, to ensure a pure sample of events whose background are only cosmic ray-induced neutrons, a minimum of 5 observed neutrons are required. 

\section{Neutron Detector System Modeling}

The system-wide single neutron detection efficiency was calibrated using a $^{252}$Cf source  placed at the center of the target and found to be 23.2\%. However, the systems response to $^{252}$Cf is not the response to muon induced neutron production because of the difference in their energy spectra. To accurately estimate the detection efficiency of a produced neutron, whether produced in the lead target or surrounding rock, a Geant4 model detector was built matching the geometry of the lead target, the polyethylene thermalizer and the arrangement of $^3$He counters. 

The $^3$He tubes detect neutrons through the reaction 
\begin{equation}
   ^{3}He + ^{1}n(thermal) \rightarrow ^{1}H + ^{3}H + Q(764 keV)
    \label{equ:N_capture}
\end{equation}

 In the Geant4 simulation, when reaction \eqref{equ:N_capture} is observed, the corresponding counter is considered to generate a signal or neutron hit, without modeling the complex ion particle motion within the gas and walls of the tubes. If the diameter of the $^{3}$He tube in the simulation is chosen to be 1.55 cm as manufactured, the average neutron efficiency for the $^{252}$Cf spectrum is calculated to be $(28.0\pm 0.1) \%$, which is larger than the measured result. The difference between the simulation predicted efficiency and the measured efficiency is due to the wall effect\cite{KnollDetector}. The wall effect reduces the detection efficiency because either the proton or the triton or both strike the detector's cathode or the non-sensitive volume at the counters' end, producing a reduced energy pulse, less than the Q in Equation \eqref{equ:N_capture}. If the energy deposition is below the set threshold, no hits are recorded. 
 
 The wall effect, that is, the reaction products $^{3}$He(n,p)$^{3}$H occurring within one mean free path length away from the wall, and the set threshold are then important parameters in calculating the system's neutron detection efficiency.  The mean free path length can be calculated analytically. Because the $^3$He tubes thresholds were set close to Q, the neutron detection efficiency is reduced by a geometric multiplying factor, 82\%, entirely due to the wall effect\cite{shalev1969wall}. Correcting the $^{252}$Cf Monte Carlo calculated calibration efficiency yields MC($^{252}$Cf)$\times$ Wall\ Effect = 0.28 $\times 0.82 \approx 0.23$, in agreement with the measured $^{252}$Cf result. 
 
 Based on these results, the method used to calculate the neutron detection efficiency was to assume a reduced active ${^3}$He diameter set to 1.33 cm and record as a hit any observation of reaction \eqref{equ:N_capture}. In order to keep the amount of polyethylene the same, the diameter of the tube is kept at 1.55 cm, but the tube has an empty gap to account for the wall effect and threshold setting.

\begin{figure*}[t]

\centering
    \includegraphics[width=0.96\textwidth]{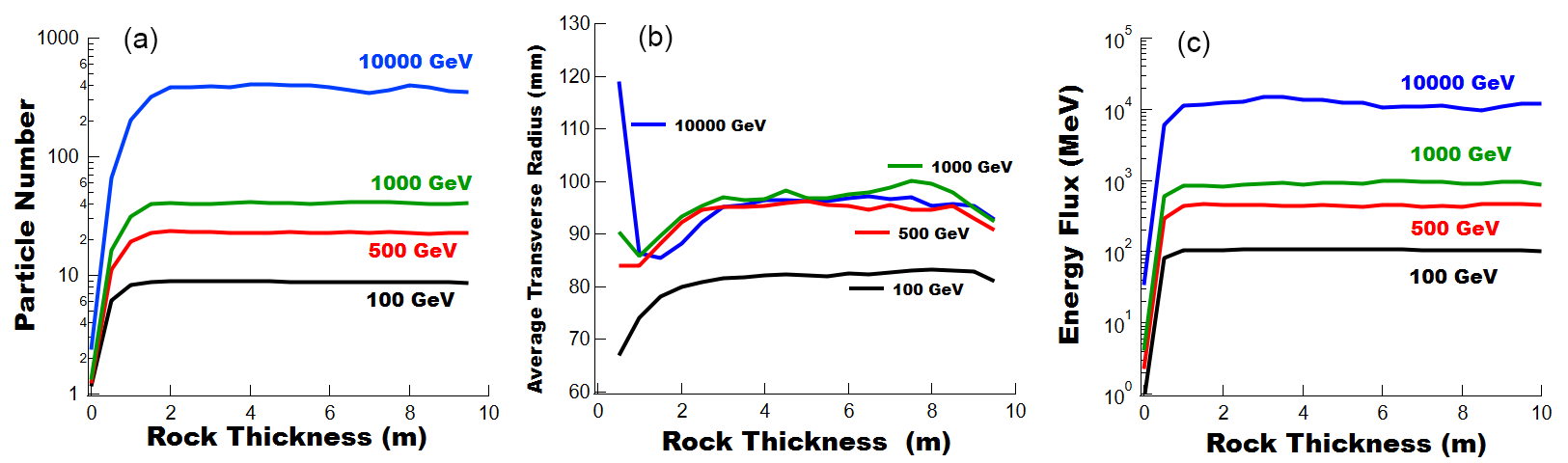}

\caption{  Simulated, (a) averaged secondary particle flux, (b) averaged secondary particle transverse radius and (c) averaged secondary energy flux passing through transverse planes as a function of muon path length.}
\label{fig:Spectrum_Three}
\end{figure*}

\section{Muon Propagation to Depth}

At experimental depth the only remaining cosmic ray particles left are muons and neutrinos. The rate of neutrino induced events is at least 6 orders of magnitude less than muon rates. The neutron production in the lead target is the result of direct interactions of the cosmic-ray muons with the target and secondary particle production by muons with the surrounding rock.  Sea-level muons are propagated to a rock layer above the detector using Geant4 to find the muon energy spectrum and angular distribution at depth. The simulation spectrum are compared to Miyake's phenomenological estimation and to experimental measurement made at the mine.

\subsection{Muon Shower Equilibrium}

Cosmic ray muons are capable of penetrating to significant depths underground. At depth, these muon produce a myriad of particles, including photons, electrons, and hadrons. 
Other cosmic ray particles and those produced as the muon passes through the rock can travel only a short distance, $\sim$ 2 m, in the rock. This short path length allows for a simplification of the neutron production modeling because the full simulation does not need to start at sea level, but only a few meters above the detectors laboratory hall.

To find the path length after which the muon-induced showers come into equilibrium, muons with energy between 100 GeV to 10 TeV are propagated through standard rock composition but having density 2.85 g/cm$^3$. The reported observables are in the transverse plane to the muon's initial momentum as a function of the distance from the muon's entry into a 10 m thick rock layer.

Figure \ref{fig:Spectrum_Three} shows the results of the Geant4 simulation. All the observables are evaluated on transverse planes to the muon's path. For all distributions, an equilibrium state is achieved within $\sim$ 2 m of rock in all ranges of muon energy. In Figure \ref{fig:Spectrum_Three} (b), the distribution is not zero when the thickness is zero. This effect occurs due to back-scatter from rock deeper along the muon's path. 

While secondary particle production quickly comes into equilibrium there is a concern that secondary muons may have a longer attenuation length than the equilibrium length of the dominating hadronic and electromagnetic components.  If so, it would cause an additional muon flux on the Pb-target beyond the sea level component propagated to depth.  If a 2.5 GeV cutoff is used, equivalent to the minimum ionizing dE/dx energy loss of a muon passing through 4m of rock, it is found that secondary muons produced by primary muons with 100 GeV, 1 TeV, and 10 TeV, having a path length greater than 4m is less than 10$^{-5}$, 10$^{-4}$, 10$^{-3}$, allowing this secondary source of neutron production in the Pb-target to be neglected. 

From these consideration it is concluded to start the full muon shower simulation in a horizontal plane 4 m above the detector's cavern hall.  Doing so captures the effects of (i) secondary particle production due to the entire mines over burden, (ii) the secondary particle contribution to neutron production in the nearby target hall rock, and (iii) secondary particle interactions within the lead target.

 \begin{figure}
\centering

\includegraphics[width=0.8\linewidth]{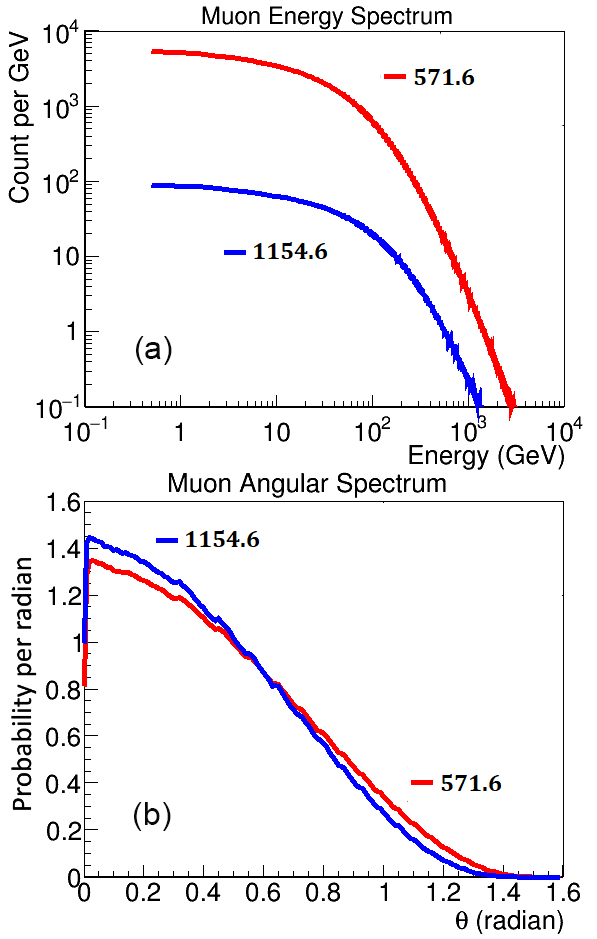}

\caption{(a) Cosmic ray muon energy  and (b) angular distributions  at 571.6 m.w.e.(6504 hr) and 1154.6 m.w.e.(1440 hr). The energy spectra are normalized to the muon number passing through a 30 cm x 30 cm square, equivalent to the size of the top of the lead target. The two angular spectra are normalized to 1.}
\label{fig:underground_spectrum}
\end{figure}

\subsection{Muon Propagation from Sea Level to Depth}
No at depth experimental or analytical correlated distributions are available. To correctly model the correlated angular and energy distribution of muons at depth, a propagation model must be used starting with the known correlated distributions at sea level \cite{CosmicRayPDG},

\begin{equation}
\begin{split}
    \frac{dN_\mu}{dE_\mu d\Omega} & \approx \frac{0.14E_\mu^{-2.7}}{cm^2\ s\ sr\ GeV}\times \\
&\left \{ \frac{1}{1+\frac{1.1E_\mu cos\theta}{115 GeV}} + \frac{0.054}{1+ \frac{1.1E_\mu cos \theta}{850 GeV}}\right \},
\end{split}
\label{equ:Surfcosmic}    
\end{equation}
and propagating individual muons to depth.

The validity of the Geant4 propagation model will be checked by (1) using Miyake's empirical model to compare separately the muon flux density and angular distributions as a function of depth and (2) using the mine muon intensity measurements as a function of depth. The depth in m.w.e. is obtained from a CUPP group's unpublished simulation model. 

Formula \eqref{equ:Surfcosmic} is valid when muon decay is negligible ($E_\mu > $100/cos\hspace{0.05cm}$\theta$ GeV) and the curvature of the Earth can be neglected ($\theta < 70^\circ$). Fortunately, the two limitations do not influence the muon spectrum at 583 m.w.e. and 1166 m.w.e. underground.  As is discussed in Appendix \ref{app:muon_propogation}, muons not suitable for formula \eqref{equ:Surfcosmic} will not pass through the thick  583 m.w.e. rock layer.

 In the Geant4 model, muons are scattered in the standard rock, lose energy, create shower particles, and can suffer catastrophic energy loss. Standard rock with density 2.65 $g\ cm^{-3}$ is used in this cosmic ray propagation model because the CUPP reported depth in m.w.e. can be directly converted to the depth in standard rock by definition.  The muons are propagated from sea level to 4 m above the laboratory. Only muons are tracked to save simulation time. Shower particles are not tracked once generated. Finally, at depth, the muon's new energy (E'), and angles ($\theta '$ and $\phi '$) are recorded, having the required energy-angular correlation. The generated underground energy and angular spectra are shown in Figure \ref{fig:underground_spectrum}.  The mean muon energy at 571.6 m.w.e. is 95.6 GeV and at 1154.6 m.w.e. is 158.7 GeV. The total muon induced event number during the experiment at 571.6 m.w.e. is $\sim 36$ times that collected at 1154.6 m.w.e.. 
 
\begin{figure}
\centering

\includegraphics[width=0.85\linewidth]{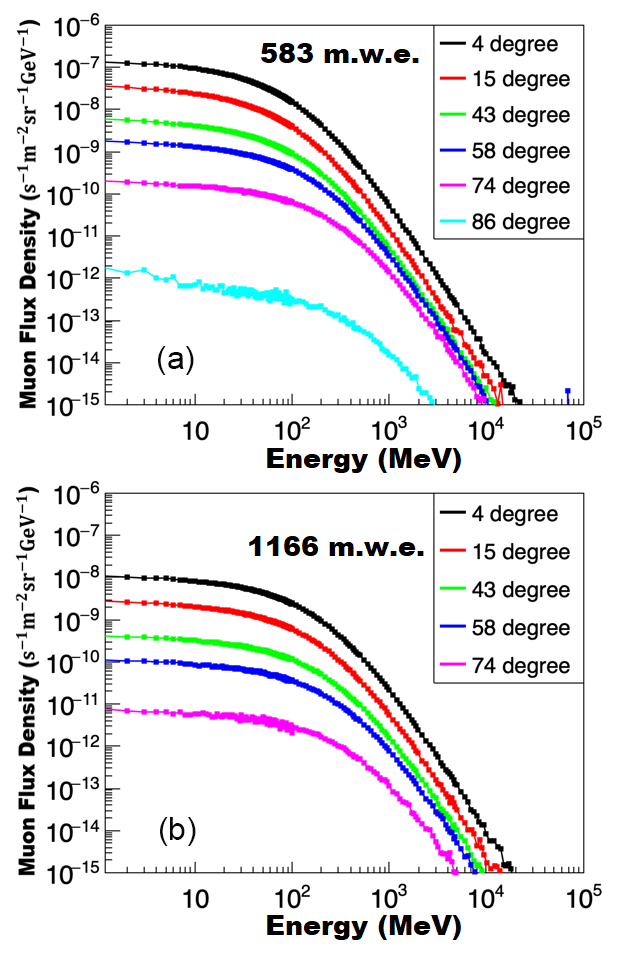}

\caption{Muon flux density as a function of angle and incident energy at the experimental depths (a) 583 m.w.e. and (b) 1166 m.w.e..}
\label{fig:Underground_spectrum_angle}
\end{figure}

 Figure \ref{fig:Underground_spectrum_angle} shows the Geant4 calculated muon energy flux density at depth as a function of angle for the two locations of the detector 4m above the detector halls.

\subsection{Geant4 and Miyake Formula Comparison }  \label{sec:cosmic_miyake}

The underground intensity of muons can be calculated with the widely used empirical formula of Miyake.  At intermediate depths (100 m.w.e. to 4000 m.w.e.), the Miyake formula is given by \cite{kf2001cosmic} 

\begin{equation}
  I(X) = \frac{A}{X+400}(X+10)^{-1.53}e^{-8.0\times 10^{-4}X}, 
\label{equ:depth2}
\end{equation}

\noindent where, \textbf{A} is the only free parameter and \textbf{X} is in m.w.e..  \textbf{A} was found by fitting measurements made at the Pyhas\"almi mine above and below the experiment\cite{enqvist2005measurements}. The flux density measurements are presented in Appendix \ref{app:CUPP_depth}. The fit returned $\mathnormal{A}=(2.97 \pm 0.114) \times 10^6~ (m.w.e.)~ m^{-2} s^{-1}$. 

\begin{figure}[H]
	\centering
	\includegraphics[width=0.85\linewidth]{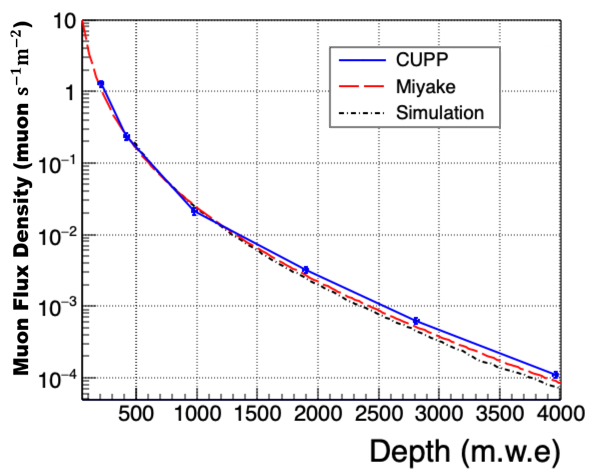}
	\caption{Muon flux density as a function of depth comparison of CUPP data points(blue), Geant4(black), and  Miyake fitted CUPP data(red).
 }
	\label{fig:muon_density}
\end{figure}

The muon cosmic ray flux density generated by the Geant4 Model and predicted by Miyake's formula along with the measurements are compared as a function of depth in Figure \ref{fig:muon_density}. The muon flux normalization at 4 m rock above the experimental cavities are compared in Table \ref{ta:density_depth}. Also shown is  the experimental underground muon cosmic ray density at 571.6 m.w.e. and 1154.6 m.w.e. extrapolated between measured points using a 5 free parameters version of the Miyake formula.

\begin{equation}
  I(X) = \frac{P1}{X+P2}(X+P3)^{-P4}e^{-P5\times X}.
\label{equ:depth}
\end{equation}

The comparison yields an average difference of $5.8\%$ over the range, 500 m.w.e. to 1200 m.w.e.. 

\begin{table}[]

\end{table}

\begin{table}
\centering
\caption{Comparison of the Geant4 and Miiyake single parameter fitted Underground muon density, 4m rock above the experimental caverns, units in $muon\ s^{-1} m^{-2}$.}

\begin{tabular}{l c c c}
\begin{tabular}[c]{@{}l@{}}Depth~~\\(m.w.e.)~~\end{tabular} & \begin{tabular}[c]{@{}c@{}}~~Miyake~~\\ ~~Fitting~~\end{tabular} & \begin{tabular}[c]{@{}c@{}}~~Geant4~~\\ ~~Simulation~~\end{tabular} & \begin{tabular}[c]{@{}c@{}}~~5-Param~~\\ ~~Fitting~~\end{tabular} \\ \hline
571.6                                                    & 0.114                                                    & 0.123                                                       & 0.105                                                     \\
1154.6                                                   & 0.0154                                                   & 0.0148                                                      & 0.0148                                                    \\ \hline
\end{tabular}
\label{ta:density_depth}
\end{table}

\begin{figure}[!ht]
	\centering
	\includegraphics[width=1.0\linewidth]{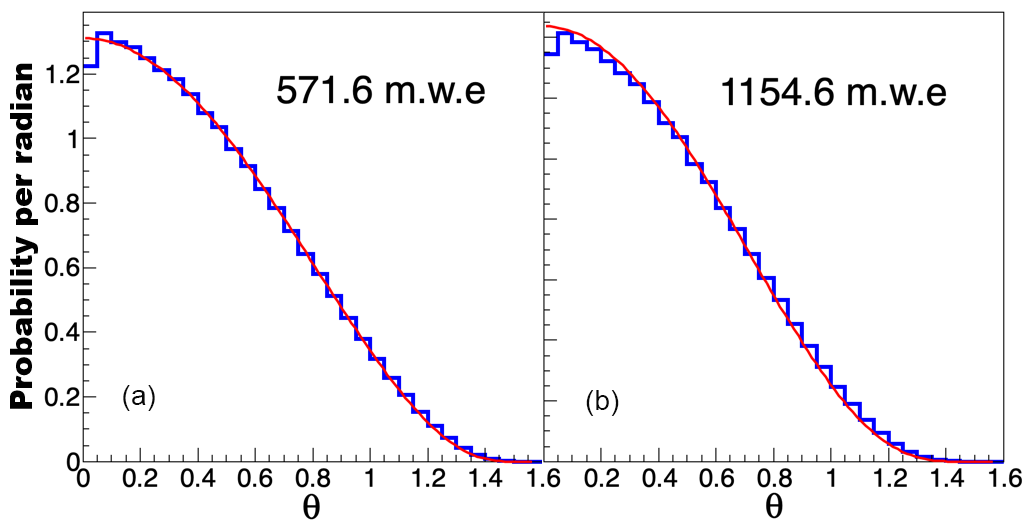}
        \caption{Comparison of the normalized cosmic ray muon angular distributions at 4 m rock above the two experimental  halls (a) 571.6 m.w.e. and (b) 1154.6 m.w.e..  Geant4 (blue).  Miyake's formula (red).}
	\label{fig:Angular}
\end{figure}

The muon angular distribution formula of Miyake\cite{kf2001cosmic} is given by, 
\begin{equation}
  I(X,\theta) = I(X,0^{\circ })cos^{1.53}(\theta)e^{-8.0\times10^{-4} X(sec(\theta)-1)}
  \label{equ:angle_dis}
\end{equation}
\noindent where $I(X,0^{\circ })$ is the vertical intensity at depth X. The underground muon angular distributions predicted by the Monte Carlo and Miyake's formula are compared at the two experimental depths in Figure \ref{fig:Angular}. At the two experimental locations the averaged rms difference between the two distributions is less than 1\%.

\subsection{Charge Ratio of the Surface Muons}
The final input to the Monte Carlo simulation is the charged muon flux ratio. At depth, for low energy muons, E$_{\mu}<$1 GeV, the  $\mu^-$ cross section with lead is larger than the $\mu^+$ due to attraction with the nucleus.  For higher energy muons, E$_{\mu}>$1 GeV, the cross sections is nearly charge independent.   This effect is displayed in Table \ref{ta:pm_compare}.  The average generated neutron number and the reaction probability for muons incident on a 30 cm cube lead target is also shown. The reaction probability is the probability of  generating at least one neutron.

To correctly take into account the effects of the muon flux charge ratio, the $\mu^+$ and $\mu^-$ Geant4 simulation are separately modeled. Only at the end of the complete simulation are they recombined using a single charge ratio value.  At 583 m.w.e. depth,  muons $<1$ GeV require a minimum energy of $\sim 150$ GeV at sea level to propagate to this level. Likewise at 1166 m.w.e. depth, an energy of $\sim 340$ GeV is required. Because the $\mu^+$ to $\mu^-$ ratio at the surface between 100 GeV and 500 GeV is constant, the value $1.29 \pm 0.13$\cite{khachatryan2010measurement} is selected. The estimated charge ratio for higher energy muons is less than the error of the selected value.

\begin{table}
\centering
\caption{The average number of produced neutrons excluding zero neutron production and the probability of generating at least one neutron for a $\mu^+$ or a $\mu^-$ incident on a 30 cm cube lead target. From the Geant4 simulation. }
\begin{tabular}{ l l c c c }

 &~~ Energy & ~~100 MeV   & 1 GeV    & 10 GeV    \\ \hline \\
\multirow{2}{*}{$\mu^-$}~~~ &  $<Neutron>$ & 4.9   & 4.7   & 6.2 \\[3pt]
    & \begin{tabular}[c]{@{}l@{}}~~~Reaction \\ ~~~Probability \\[6pt]\end{tabular}         & 0.87     & 0.0013  & 0.012  \\[6pt]
\multirow{2}{*}{$\mu^+$} & $<Neutron>$ & 1.4      & 4.3     & 6.5     \\[3pt] 
    & \begin{tabular}[c]{@{}l@{}}~~~Reaction \\ ~~~Probability\end{tabular}  & 0.0034  & 0.0014  & 0.011  \\ \hline
\end{tabular}
\label{ta:pm_compare}
\end{table}

\section{Neutron Production Simulation}

\subsection{Geant4 Model}

All muon and secondary particle processes in the rock and lead are simulated with Geant4-11.01 using the physics list QGSP-BERT-HP.  This list was selected based on its use for LHC experiments. In particular ATLAS and CMS have studied the physics performances of the physics lists and converged on the use of the QGSP-BERT physics list as the most comprehensive and thus the default \cite{kiryunin2006geant4}.

\subsection{Model Simulation Universe}

The Geant4 simulation begins full shower modeling starting in a horizontal layer 4m above the laboratory hall.  Muon generated showers are fully developed allowing the shower particles to interact with the lead target and detector system to generate neutrons. 

\begin{figure}[!ht]
\centering

\includegraphics[width=1.0\linewidth]{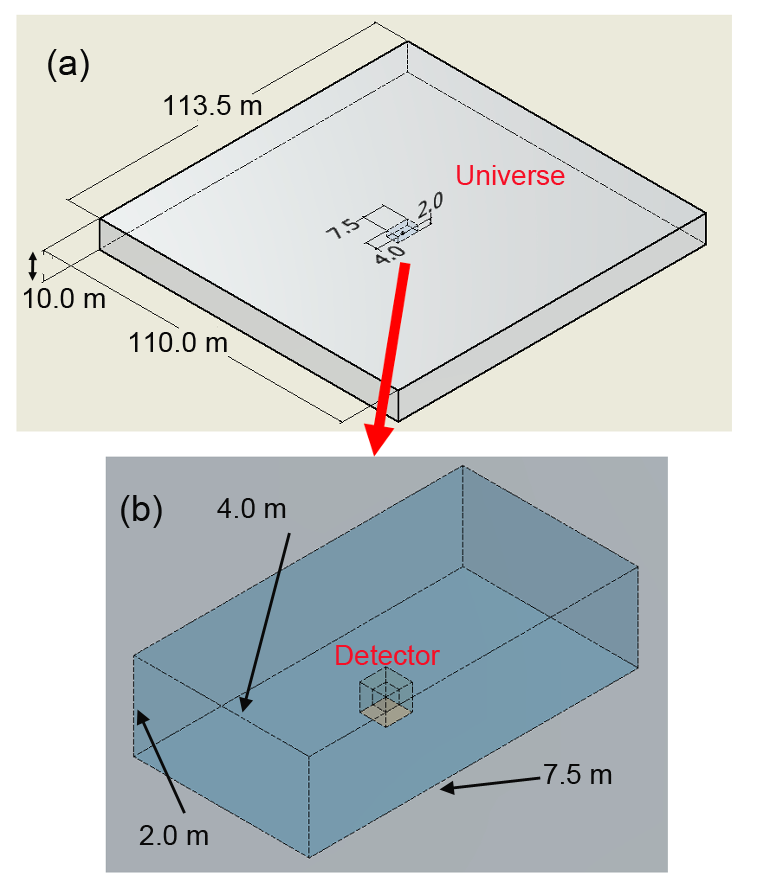}

\caption{(a) Proportional illustration of the Geant4 Universe, includes the rock layer and (b) the laboratory cavern and the detector system, 30 cm Pb-cube inside 60 cm poly-cube sitting centered on the cavern floor.}
\label{fig:Model_geo}
\end{figure}

The model universe is displayed in Figure \ref{fig:Model_geo}. The NMDS-II detector system, is centered on the floor inside a cavity ($7.5 m \times 4 m \times 2 m$) surrounded by rock. The horizontal thickness of each of the universe's rock side walls is 53 meters.  The cavity rests on and is capped by a 4m thick rock floor and  roof. Both extend to the outer edges of the side walls forming a rectangular cube. The dimensions were chosen to assure cosmic ray muons with angles from 0 degree to 85.5 degree intersect the top of the universe's roof. The number of slant angle muons passing through the target not covered by this angular range is smaller than 1 over the course of the experiment. 
The Geant4 simulation inputs are then, (1) the muon flux intensity at depth, (2) the correlated muon energy and angular distribution at depth and finally, (3) the cosmic ray muon charge ratio.
The full simulation starting point is the universe's top surface or roof, $113.5 m \times 110 m$. The total number of muon events intersecting the universe's roof in the experiments live times are shown in Table \ref{ta:Normalization}. The values are found using Eq.~\ref{equ:depth}, Miyake's fit to the experimentally measured \cite{enqvist2005measurements} muon flux density values, shown in Table \ref{ta:density_depth}.   
In order that the simulations have superior statistics compared to the data, more than 10 times the experimental statistics were simulated, also displayed in Table \ref{ta:Normalization}.

\begin{table}
\centering
\caption{Number of cosmic ray muons intersecting the roof of the Geant4 universe during the experimental live times and the number of simulated events at each detector depth. \\ }

\begin{tabular}{l c c}
       Depth      & ~~583 m.w.e. & ~~1166 m.w.e. \\[3pt]\hline
Live Time hrs & 6504 & 1440\\[3pt]
$\mu$-Experiment   & ~~$3.3\times 10^{10}$ & $1.0 \times 10^8$  \\ [3pt]
$\mu$-Simulation   &~~ $4.6\times 10^{11}$ & ~$2.3\times 10^{11}$  \\ \hline

\end{tabular}
\label{ta:Normalization}
\end{table}

\begin{figure}
\centering
\includegraphics[width=0.8\linewidth]{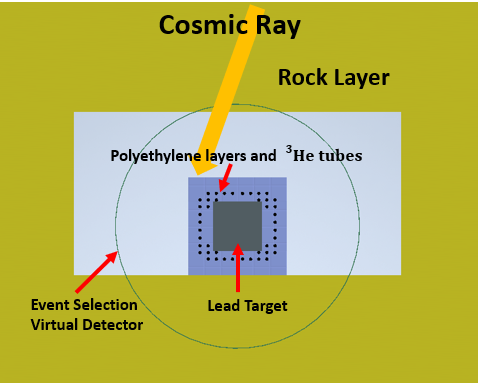}
\caption{Illustration of the Geant4 Event Selection Virtual Detector, relative to the $^3He$ detector system and Pb-target centered on the cavern floor. The dimensions are not in proportion.}
\label{fig:Model_illustration}
\end{figure}

\subsection{Full Event Simulation Trigger}

Because of the large size of the simulation universe and the corresponding low interaction rate of cosmic ray showers with the experiment's detectors, a simulation trigger was developed to further reduce the required computational run time. Once the simulation located a muon on the roof of the universe, using its initial angular parameters the muon track was pointed at the target. The intersection of the muon track with a spherical surface virtual detector was checked and the event was rejected from further consideration if not intersecting the virtual detector. This virtual detector forms a MC trigger for which muons will be fully simulated as they pass through or nearby the target. Figure \ref{fig:Model_illustration} illustrates the geometry of the model.

For an accurate simulation, it is critical that the event selection virtual detector should be given an appropriate size. Otherwise,  some events with neutrons detected will be mis-rejected if the size is too small. However, computing time is roughly proportional to the volume of the virtual detector, or $R^3$. Too much computing time will be taken if the event selection virtual detector is too large.

To optimize the radius of the virtual trigger detector, simulations were performed to study the number of detected neutrons as a function of the virtual detector radius, ranging from 30 cm to 360 cm. In the region where the neutron multiplicity exceeds 4, all curves with a radius R $\geq$ 300 cm are indistinguishable using 10 times experimental statistics, while those with R $<$ 300 cm remain distinct from one another.
Based on these results, a virtual trigger detector radius of R = 300 cm was selected as the optimal value.

\subsection{Simulation Statistical Errors}

Finally, once a muon is accepted by the virtual trigger, a full shower simulation is completed. In order to get the average response of the NMDS-II $^3$He neutron detector system, the neutrons hit times on the $^3$He tubes were re-sampled 100 times. The statistical errors for the simulations detected neutron multiplicity are estimated using a Bootstrapping method\cite{johnson2001introduction}.
\section{Experimatal and Simualtion Data Analysis}

\subsection{The Mechanism of Neutron Multiplicity's Power Law Distribution}

When a high-energy muon and/or its associated shower particles enters the  NMDS-II lead target, a neutron cascade can be initiated. The Pb-target can be thought of as a many-body quantum system with the initial cascade particles having been promoted to a highly excited quantum state. As the cascade or avalanche progresses, the lower-energy quantum states are excited until neutron production is exhausted. After each neutron emission, the number of possible remaining transitions decreases, making the process
irreversible and reducing the available quantum states.

This cascade  process closely matches a sample-space-reducing (SSR) process, where each step in the cascade narrows the system’s future possible states. 
Importantly, SSR processes are known to naturally generate power-law distributions, $y \propto n^{-p}$ \cite{doi:10.1073/pnas.1420946112},  with the index remaining independent of nuclear structure and reaction type\cite{fujii2021power}. In the case of muon-induced neutron cascades, this means the distribution of the number of emitted neutrons or the energy released is expected to follow a power law. Many exponents for cascades or avalanche-like processes are found within a range of $p \in (0,3) $. where p represents the multiplicity factor of the average transition probability at each step in the cascade. 

The relationship between cascade processes and universal power-law distributions is based on two fundamental principles: the stochastic nature of transition rates and the exponential energy dependence of the level density in fermionic many-body systems\cite{fujii2021power}. It is also proved that the exponent index p should be 2 if the energy is conserved\cite{corominas2017sample, fujii2021power}. However, in the case of neutron generation in lead target, the energy deposited in the lead target is not necessarily released solely through neutrons. As a result, the spectrum may decay more rapidly, implying that $p>2$. For instance, the exponent parameter value is 2.4 in solar neutron emissivity \cite{powerlaw_Solar}.

Furthermore, a power-law dependence has also been observed in other physics processes, such as $\gamma$-ray intensity during heavy nuclei transitions\cite{fujii2021power}, as well as the line intensity distribution of many-electron atoms in plasmas\cite{BAUCHEARNOULT1997441}.

\subsection{NMDS-II Simulation and Experimental Data Power Law Fitting}

\begin{figure}[!ht]
\centering
\includegraphics[width=0.95\linewidth]{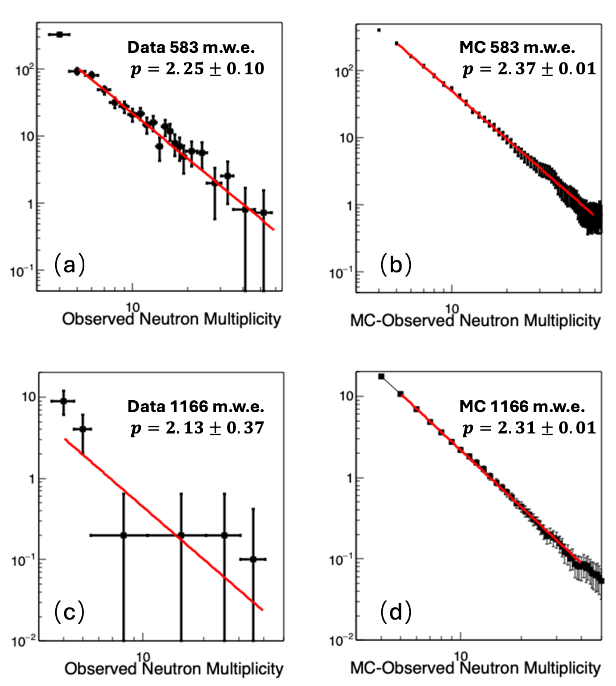}
\caption{The neutron multiplicity distributions (black) with  power law fit(red).  (a) 583 m.w.e. data. (b) 583 m.w.e. simulation. (c) 1166 m.w.e. data. (d) 1166 m.w.e. simulation.}
\label{fig:p_values}
\end{figure}

The MNDS-II neutron multiplicity experimental data as well as Geant4 simulations results are independently displayed in Figure \ref{fig:p_values}. These data are indepentently fit using a two parameter power law, $y = k\times n^{-p}$. These fits are compared with data and used later for the dark matter search to account for the cosmic ray background.

At 583 m.w.e., a multiplicative 5.6\% systematic error at each experimental point was included and  placed in quadrature with the statistical error.  The additional error is due to deterioration of the $^3$He tube over the course of data collection. As the case with the simulation, the data fit well to a power law yielding $\chi^2$/DoF = 0.72 for 54 degrees of freedom.  

Likewise, the Geant4 simulation results fit a power-law well with $\chi^2$/DoF = 1.24 at 583 m.w.e.  and $\chi^2$/DoF = 1.24 for the 1166 m.w.e..

For the data at 1166 m.w.e. a Maximum Likelihood Estimation (MLE) method is used to measure the power law parameters for the data whose neutron multiplicity is equal to or larger than 5 because there are only 6 points in the data set. In addition, the multiplicity probability distribution is normalized to one, yielding a single-free parameter fit for the index p. The fit then yields p=2.13 with an only slightly asymmetric Gaussian shaped Likelihood distribution having  $\sigma_{RMS} = $  0.37. 

The fitting and MLE results for NMDS-II experimental data and Geant4 simulation are summarized in Table \ref{ta:fit_para}.  As expected for SSR processes, the index value p, within errors, are in agreement for the Geant4 Monte Carlo simulations and the data. Likewise, we expect the p value to be close to 2.4, as obtained from solar the neutron spectrum.

\subsection{Errors in the Simulation Model}
The Geant4 simulation agrees with the experimental data in the exponent parameter p but not in the amplitude parameter k.

The difference in k parameter between the Geant4 simulation and experimental data is mainly caused by the systemaic error in the simulation, such as the error of muon-lead cross section. However, even if there are systematic errors in the simulations, it is not surprising that the simulations and experimental data agree in p due to the characteristics of the SSR process.

The difference in k parameter or the amplitude between the Geant4 and FLUKA simulations is a measure of their systematic error. Muon mean energy at 583 m.w.e. is $\sim$ 95 GeV, with corresponding mean neutron production ratio Geant4/FLUKA $\sim 1.3$ and  muon mean energy at 1166 m.w.e. is $\sim$ 160 GeV, with corresponding ratio Geant4/FLUKA $\sim 1.4$. This systematic error in mean neutron number produces a larger error in the amplitude parameter. This effect can be illustrated by scaling the neutron production by a factor $\alpha$, which leaves the power law invariant, but significantly affects the amplitude, as shown in the formula \eqref{eq:K},

 \begin{equation}
     k\times(n\alpha)^{-p}=k\alpha^{-p}\times n^{-p}=k'n^{-p}.
     \label{eq:K}
 \end{equation}

This shows that the amplitude parameter k is too sensitive to systematic errors in the simulation, making the comparison between the data and the simulation in this parameter unavailing.

\begin{table}
\centering
\caption{ Geant4 simulation and  NMDS-II data, neutron multiplicity distribution  power law function, $k\times n^{-p}$, fit parameters. The fits are over the range [5,60] observed neutrons. All p parameters can be compared. } 

\begin{tabular}{l c l c c c}

\begin{tabular}[c]{@{}l@{}}Depth\\ (m.w.e.)\end{tabular} & \begin{tabular}[c]{@{}c@{}}Time\\ (hrs)\end{tabular} &            & p                                                            & \begin{tabular}[c]{@{}c@{}}$\chi^2$ \\per DoF\end{tabular} \\ \hline
\multirow{2}{*}{583}       & \multirow{2}{*}{6504}       & Geant & \begin{tabular}[c]{@{}c@{}}2.37\\ $\pm$ 0.01\end{tabular}    & 1.24             \\ \cline{3-5} &              & Exper & \begin{tabular}[c]{@{}c@{}}2.25 \\ $\pm$ 0.10\end{tabular}  & 0.72   \\ \hline
\multirow{2}{*}{1166}      & \multirow{2}{*}{1440}       & Geant & \begin{tabular}[c]{@{}c@{}}2.31 \\ $\pm$ 0.01\end{tabular} & 1.24      
\\ \cline{3-5} 
&    & Exper & \begin{tabular}[c]{@{}c@{}}2.13\\ $\pm$ 0.37\end{tabular}  & 0.76                                                       \\ \hline
\end{tabular}
\label{ta:fit_para}
\end{table}

\begin{figure}
	\centering
	\includegraphics[width=0.9\linewidth]{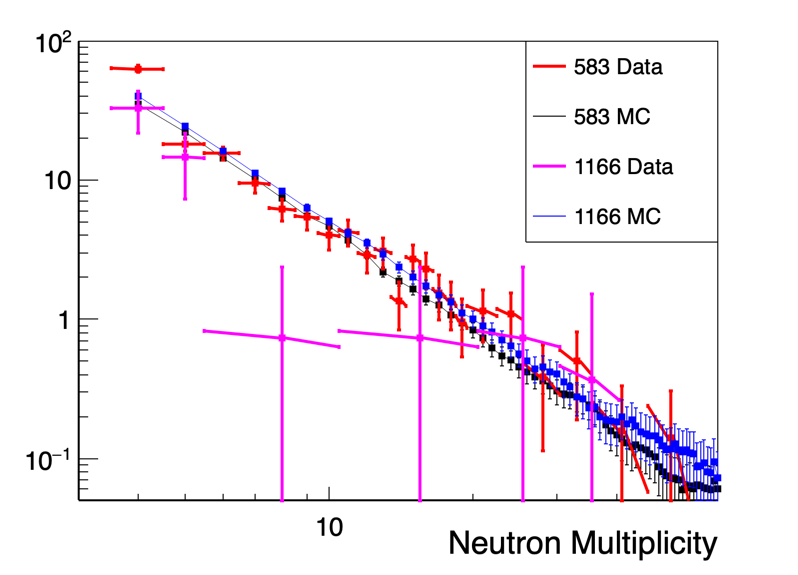}
	\caption{Amplitude independent comparison of the simulations and experimental neutron multiplicity distributions at both depths. Each curve is fit using $k\times n^{-p}$, however k is shifted to 1000, an arbitrary number, for all distributions for comparison of the power law slope parameter p-values.}
	\label{fig:NoK}
\end{figure}

The distribution of neutron multiplicity for the data sets and simulations is shown independently of k in Figure \ref{fig:NoK} by fixing k at an arbitrary value. If the power law index p changes as a function of depth, it is a slow change.

\section{Dark Matter Search Method}
 Because it is assumed that DM-M cross sections are extremely small, the dark matter flux is expected to be independent of depth.  However, the sensitivity to DM neutron production is not the same at both depths.  The cosmic ray flux is reduced by a factor of $\sim 7.5$ at the 1166 m.w.e level compared to the 583 m.w.e. depth. In addition the data collection time was 4.5 times longer at the 583 m.w.e. level. Even so, if all the events collected at the 1166 m.w.e. level are assumed to be DM induced events, only  $\sim 3$\% of the events at the 583 m.w.e. level would be DM matter induced events.

Due to the much larger background at the 583 m.w.e. level the dark matter-matter search is based on neutron multiplicity event data collected at 1166 m.w.e.. 
 
 In this dark matter search, it is assumed that the cosmic ray neutron multiplicity distributions have a common  power law index. A normalized probability function is produced consisting of the combination of the cosmic ray induced neutron production spectrum, $k\times n^{-p}$ and a DM hadronic neutron production model spectrum $S_{DM}$
 \begin{equation}
     \mathcal{P}(\beta;n)=(1-\beta)k\times n^{-p}+\beta S_{DM}(n).
     \label{eq:prob}
 \end{equation}
 From this the likelihood function is formed,
 \begin{equation}
     \mathcal{L}(\beta)=\Pi^{n=60}_{n=5}~\mathcal{P}(\beta;n)^{\mathcal{D}_(n)},
     \label{eq:likely}
 \end{equation}
 where $\mathcal{D}$ is the data spectrum at 1166 m.w.e. and the p parameter is set to the simulation value at 1166 m.w.e. found in Table \ref{ta:fit_para}.

The significance $\theta$, of any DM signal $S_{sig}$,  in standard deviations is,
\begin{equation}
    \theta = \frac{S_{sig}}{\sqrt{S_{sig}+B_{back}}}=\frac{\beta_{peak} \times N_{1166}}{\sqrt{N_{1166}}}=\beta_{peak}\sqrt{N_{1166}}
    \label{eq:significance}
\end{equation}
where $B_{back}$ is the cosmic ray induced background, $\beta_{peak}$ is the most likely model dependent $\beta$ value of the likelihood function, and $N_{1166}=6$ is the total number of events in the neutron multiplicity spectrum at 1166 m.w.e..

\section{Dark Matter Models}

In this search, for simplicity it is assumed the complete dark matter particle's mass-energy is deposited hadronicly in the target. For example, a neutrino can transfer a fraction of its mass energy into hadronic energy. In this case a differing model depositing an on-average-fractional energy $f_{average}$ merely causes a shift in the DM converted energy in these simulations by an amount 
\begin{equation}
    E_{converted}=f_{average} ~\times ~M_{DM}.
    \label{eq:fraction}
\end{equation}

 Two extreme models are used to simulate the interaction process. A spallation model in which the DM mass energy is converted into the kinetic energy of a single proton and a pion fireball model in which the net momentum is zero.  Geant4 is used to propagate the secondary particles,  create neutrons in the target, and then follows each neutron for detection in the $^3He$ tube. In the acceptance calculation it is assumed dark matter - matter interactions happen at a random location inside the lead target.

 In both the simulations and the experimental data analysis a minimum of 5 observed neutrons are required in order to assure that radioactivity in the surround rock is eliminated as a source or cause of an event.
 
Unlike neutron production from high-energy muon interactions with lead, hadronic neutron production from proton–lead or pion–lead interactions is better understood, allowing for more precise simulations. For validation the spallation model simulation was compered to the NESSI(Neutron Scintillator and Silicon Detector) experimental results\cite{filges2001spallation}\cite{letourneau2000neutron} in the range 1.2 GeV to 2.5 GeV. The experimental neutron production and the simulation are in agreement at the 10\% level\cite{Cao2023}.

\subsection{Spallation Model}
The process of dark matter energy deposition in the lead target is modeled using a proton induced spallation model to break apart the target nuclei. In the model, the proton's kinetic energy is set equal to the mass-energy of the dark matter particle. The dark matter events are uniformly distributed throughout the lead target, and the protons' momentum vectors are selected to be isotropic at their creation point.  In the simulation neutron production is measured by the creation of an $^3H$ particle in the $^3He$ tube.

\begin{figure}[!ht]
	\centering
	\includegraphics[width=1.0\linewidth]{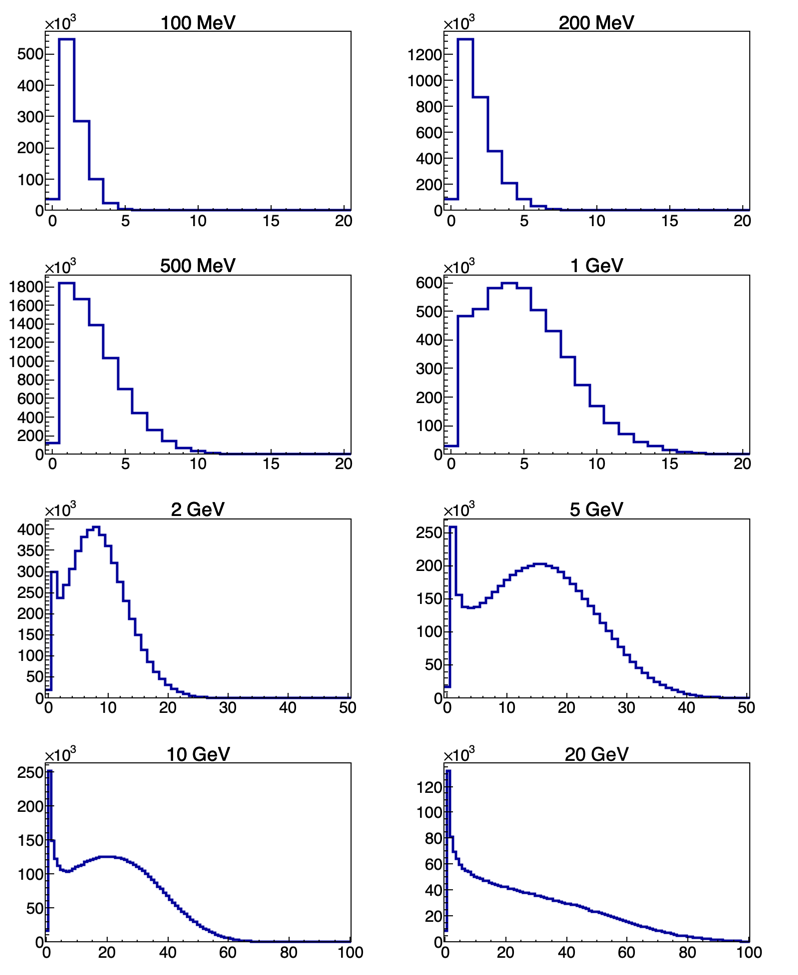}
	\caption{Unnormalized detected neutron multiplicity event distribution at a fixed deposited DM mass-energy using the spallation model. }
	\label{fig:Spallation_signal}
\end{figure}

\begin{figure}[!ht]
	\centering
	\includegraphics[width=0.8\linewidth]{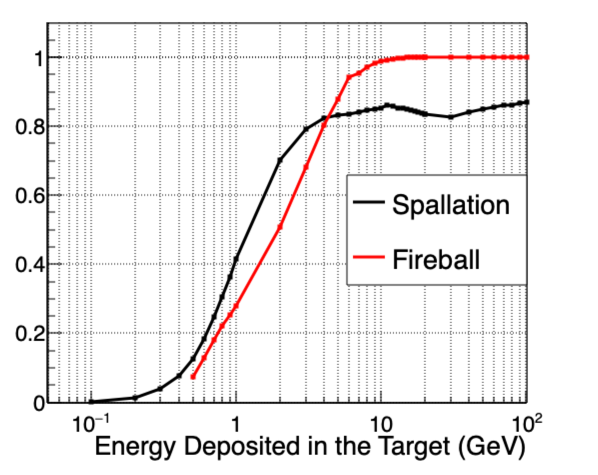}
	\caption{The model-dependent acceptance for DM-M interactions in the NMDS-II Detector.}
	\label{fig:two_Acceptance}
\end{figure}

For the DM simulation the resulting unnormalized neutron multiplicity events distributions are displayed in Figure \ref{fig:Spallation_signal} as a function of a fixed converted DM mass-energy. From these simulations the acceptance for the spallation model is presented in Figure \ref{fig:two_Acceptance}. In this model the acceptance does not plateau at 1 because near the edges of the target the momentum of the proton can point away from the target reducing the probability of an acceptable event. 

\subsection{Fireball Model}
A hadronic fireball is an intermediate state created immediately after a collision or an annihilation like process. To contrast it with the spallation model the fireball is created at rest composed completely of pions as the initial secondary particles. In the simulation charge is conserved in the creation of pions($\pi ^0, \pi ^+, \pi ^-$). 

The pion kinetic energy, E distribution is assumed to follow the Planck distribution, 
\begin{equation}
 \frac{dN}{N} = A \times \frac{E^{2}dE}{e^{E/kT}-1}
 \label{equ:fireball}
\end{equation}
where dN represents the number of particles in the energy range E to E + dE, A is the 
normalization coefficient and kT = 0.165 GeV \cite{saha1987energy}. 

 In the acceptance calculation all the pions of a single event are generated at the same point. The events are uniformly distributed throughout the lead target.  While each pion momentum vector is selected isotropically, the total momentum of an event is not required to be conserved.  Geant4 is used to propagate the pions in the lead, creating and tracking secondary particles and finally to register the detection of neutrons in the $^3He$ tubes. 

\begin{figure}[!ht]
	\centering
\includegraphics[width=1.0\linewidth]{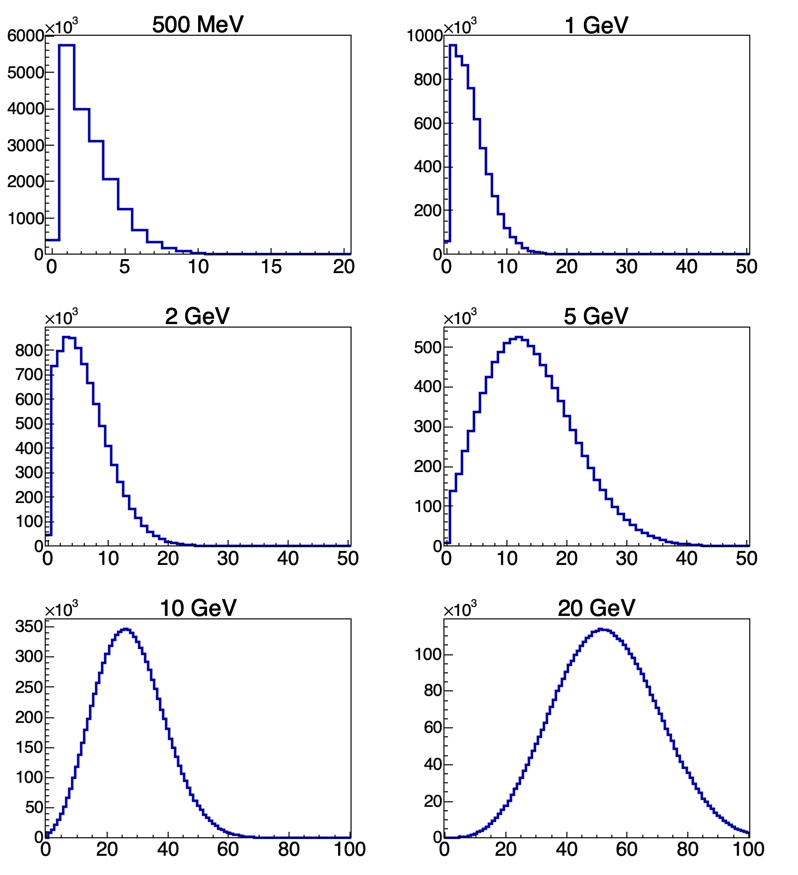}
	\caption{Unnormalized detected neutron multiplicity event distribution at a fixed deposited DM mass-energy using the fireball  model.}
	\label{fig:Fireball_signal}
\end{figure}

The resulting fireball model neutron multiplicity events distributions for a fixed DM mass-energy deposition is displayed in Figure \ref{fig:Fireball_signal}. Again the acceptance curve is displayed in Figure \ref{fig:two_Acceptance}.  The acceptance plateaus at 1 because even near the edge of the Pb-target the uniform angular distribution of the pions assures  some energy is always moving inwards.

\section{Dark Matter Signal Search}

\begin{figure}[!ht]
\centering
\includegraphics[width=0.9\linewidth]{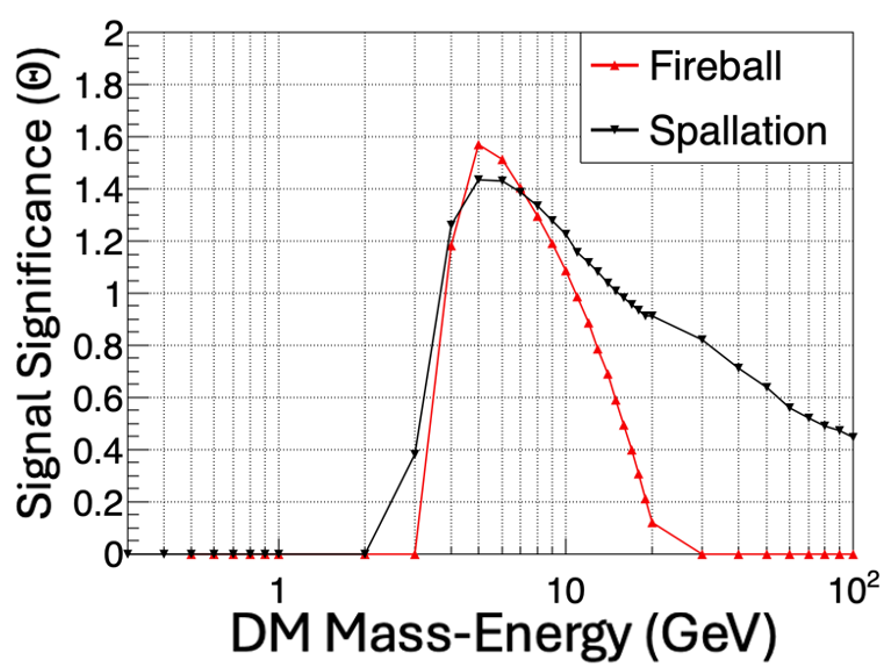}
\caption{Model dependent DM signal search as a function of the complete conversion of the DM mass-energy. Signal in standard deviations, $\theta$ verses DM converted energy.}
\label{fig:theta}
\end{figure}
The results of combining the cosmic ray background neutron multiplicity shape with the Geant4 neutron DM induced multiplicity distribution shapes through the likelihood function, Equations \ref{eq:likely} and \ref{eq:significance}, are shown in Figure \ref{fig:theta}.
The search for excess neutron multiplicity production is conducted using the data at 1166 m.w.e. displayed in Figure \ref{fig:p_values}. For the search, $\beta$ is limited to the physical positive signal region [0,1] and $\theta$ is the signal significance.

 There are mass regions in Figure \ref{fig:theta} in both models for which the most likely projection of the DM induced hadronic energy deposition yields zero in the likelihood function. This is most notable at high observed DM energy deposition above 20 GeV.  For low energy deposition less than 2 GeV the detector loses its sensitivity as the DM matter induced neutron multiplicity events distribution shape becomes more cosmic ray like. There is a peak-like shape in the signal significance between 2 GeV to 20 GeV. The top of the peak for both models is a approximately $\theta\sim$1.5, corresponding to $\sim$4 excess events. The significance is not large enough to claim a dark matter signal. 

For both models if the DM mass energy conversion to hadronic energy were not a delta function the neutron multiplicity events distribution will broaden in Figures \ref{fig:Spallation_signal} and \ref{fig:Fireball_signal}.  In this case the acceptance is reduced as well as the ability to differentiate the DM shapes from the cosmic ray background.

\section{Dark Matter  Cross Section Upper Limits}

It is assumed that DM particles are distributed uniformly and are at rest, in other words, they have no angular momentum relative to the galactic center.  The velocity of the Dark Matter flux around the earth has two components in the earth frame: the speed derived due to the sun about the galaxy and the motion of the earth relative to the sun. In this search the small variation caused by the earth's motion is discarded.
 
 The number of DM particles passing through the lead target during data collection is expressed by
 \begin{equation}
 N_{DM} = \frac{\rho_{DM}\times v \times t\times S_{target}}{m_{DM}} = \frac{6.0 \times10^{16} GeV}{m_{DM}},
 \end{equation}
where $\rho_{DM}$ is the local DM density($0.55 \pm 0.17\ GeVcm^{-3}$), v is the rotation velocity of the sun($233 \pm3\ km/s$), t is the experiment's live acquisition time (1440 hours) and $S_{target}$ is the cross section of the lead target($900\ cm^2$). 
 
 The number of expected dark matter - matter conversion events is then given by 
 \begin{equation}
     N_{events} = \frac{N_{DM}\times N_{Pb}\times\sigma_{conversion}}{S_{target}}.
     \label{eq:events}
 \end{equation}
Where $N_{Pb} = 8.9 \times10^{26}$ for the 30 cm cubic Pb-target. $\sigma_{conversion}$ is the cross section of dark matter interacting with a nucleus converting all its mass-energy into hadronic-energy. 

The model dependent 90\% upper limits, $\beta_{90\%}$ are found by integrating the normalized asymmetric likelihood function, Equation \ref{eq:likely}, over $\beta$ from zero to 90\%.

\begin{figure}[!ht]
\centering
\includegraphics[width=0.9\linewidth]{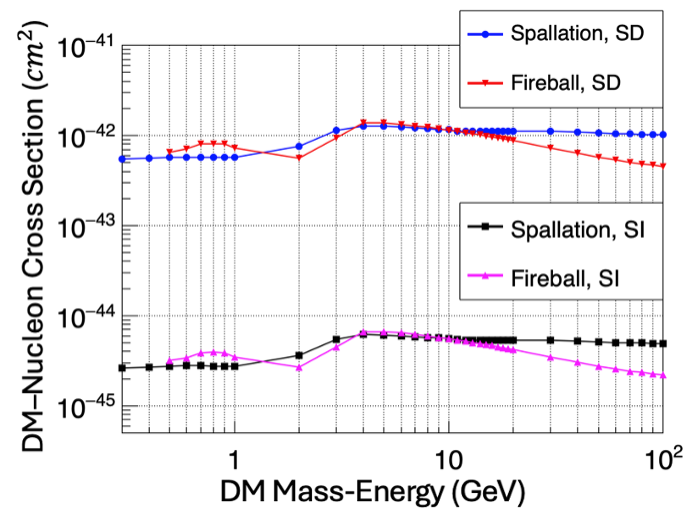}
\caption{ Model dependent, indirect detection search, 90\% cross section upper limits for DM-matter interactions using the  NMDS-II  neutron multiplicity events distribution. The results are based on 1440 hrs of data collection at 1166 m.w.e. using a 305 kg Pb-target for spin independent (SI) and spin dependent (SD) DM-matter interactions.}
\label{fig:Limits}
\end{figure}

From these results and inverting Equation \ref{eq:events}, the 90\% confidence cross section upper limits are given by

\begin{equation}
     \sigma_{90\%} = N_{1166}\frac{\beta_{90\%}\times S_{target}}{A^{2\ or\ 1} \times N_{DM}\times N_{Pb} \times \epsilon}
     \label{}
 \end{equation}
where $\epsilon$ is the model dependent acceptance and A is the atomic number.  Four limit cases are presented in Figure \ref{fig:Limits}. Spin dependent and independent limits based on the fireball and spallation models.

For spin-independent cross sections,  the sensitivity per nucleon requires dividing the factor of the atomic number $A$, twice. The first factor of A is simply due to  the target nucleon number.  The second factor occurs because the De Broglie wavelength of an at rest DM particle with mass less than 100 GeV is larger than the lead nuclear dimensions used as the target, meaning the dark matter particles interact with the nucleus as a whole. This large enhancement factor is the motivation for the use of the instrumented Pb-target.

For spin-dependent cross sections, the sensitivity per nucleon results in a single factor of A  because the interaction only take places with a single nucleon.

The  upper limits displayed in Figure \ref{fig:Limits} are based on complete conversion (a fixed delta function) of the DM mass-energy into hadronic energy.

\section*{Acknowledgement}
 The authors thank Dr. Thomas Ward from TechSource, Inc for providing the data at NMDS-II experiment and giving us useful advice in the discussion. We thank Dr. Norbert Neumeister and Purdue CMS group for allowing use of excess cluster cycles.

 This work was funded under a grant from the U.S. Department of Energy Office of Nuclear Energy, Contract No. DE-SC0007884, Department of Physics and Astronomy at Purdue University and TechSource Inc.

\appendix

\appendix

\section{Muon Energy Loss in the Rock} \label{app:muon_propogation}

 Muons lose energy by ionization and by three radioactive processes; bremsstrahlung, production of electron-position pairs and photonuclear interactions. The muon energy loss was calculated using the form \cite{CosmicRayPDG}, 
 
\begin{equation}
-\frac{dE_\mu}{dX} = a + bE_{\mu},
\label{equ:ELoss}
\end{equation}

\noindent where $\mathnormal{a}$ is the ionization loss and $\mathnormal{b}$ is the sum of the three fractional radiative losses.  Both are slowly varying functions of muon energy as shown in Table \ref{ta:MuonELoss} for standard rock.
A second-order polynomial was used to fit $\mathnormal{a}$ and $\mathnormal{b}$, each independently, as a function of $log_{10}E(GeV)$, yielding the fits values,
\begin{equation}
a = - 0.005 * (log_{10}E)^2 + 0.277*log_{10}E + 1.9
\label{equ:ELoss_a}
\end{equation}
\begin{equation}
b = - 0.1775 * (log_{10}E)^2 + 1.7105*log_{10}E + 0.3575    
\label{equ:ELoss_b}
\end{equation}
with only 1 degree of freedom.  These values are input in Equation \eqref{equ:ELoss} yielding the point-by-point transformation of the surface energy distribution to one at depth.

\begin{table}[H]       
    \centering
    \caption{Average muon range R and energy loss parameters calculated for standard rock\cite{CosmicRayPDG}. \\} 

    \scalebox{0.87}
    { 
    \begin{tabular}{|c c c c c c c c|}
    \hline
     $\mathbf E_{\mu}$ & \bf R & \bf a & $\mathbf{b_{brems}}$ & $\mathbf{b_{pair}}$ & $\mathbf{b_{nucl}}$  & $\mathbf{bE_\mu / a}$ & $\mathbf{\delta E\ water}$\\
    \hline
    \bf GeV & \bf km.w.e & $\mathbf{MeV} $ &\multicolumn{3}{|c|}{$\mathbf{10^{-6}}$} & &$\mathbf{GeV} $   \\
    & &$\mathbf{g^{-1} cm^2} $& \multicolumn{3}{|c|}{$\mathbf {g^{-1} cm^2}$} & $-$ &$\mathbf{m^{-1}} $\\
    \hline
    10    & 0.05   & 2.17   & 0.70 & 0.70 & 0.50  & 0.0088 & 0.219\\
    100   & 0.41   & 2.44   & 1.10 & 1.53 & 0.41  & 0.1246 & 0.274\\
    1000  & 2.45   & 2.68   & 1.44 & 2.07 & 0.41  & 1.463  & 0.660\\
    10000 & 6.09   & 2.93   & 1.62 & 2.27 & 0.46  & 14.85  & 4.643\\
    \hline
    \end{tabular}
    }
  \label{ta:MuonELoss}
\end{table}

 The validity of this formulation when propagated to NMDS-II depth is supported by a comparison of the formulation and measured data\cite{CosmicRayPDG}, displayed together in Table \ref{ta:MuonR}.

\begin{table}[H]                            
   \centering
    \caption{Comprision between the measured range and predicted average muon range using Formula \eqref{equ:ELoss}, units in m.w.e.. \\} 

       \scalebox{0.9}
    { 
   \begin{tabular} {c c c c} 

   ~$\mathbf E_{\mu}$ & \bf~~ Measured & \bf~~ Predicted& \bf ~~Difference\\
    \hline
    10    & 50    & 48.7  & 2.6  \% \\
    100   & 410   & 409   & 0.24 \% \\
    1000  & 2450  & 2455  & 0.21 \% \\
    10000 & 6090  & 6500  & 6.3  \% \\
    \hline
    \end{tabular}
    }
  \label{ta:MuonR}
\end{table}

 The predicted average muon range in Table \ref{ta:MuonR} is obtained by numeral calculations. The muon path in the rock is divided into many small steps. Muon energy loss at each step is then
\begin{equation}
    \Delta E =\frac{dE}{dx}\cdot  \Delta x
\end{equation}

The length of each step was taken to be $\Delta x = 0.01$ m.w.e., and $\frac{dE}{dx}$ is found from Equation \eqref{equ:ELoss}. The energy of the muon is changed in each step, so that \textbf{a} in \eqref{equ:ELoss_a} and \textbf{b} in  \eqref{equ:ELoss_b} are also changed. The predicted muon average range in Table \ref{ta:MuonR} is the total length of the muon path in the rock when the energy becomes 0.

\begin{figure}[H]
	\centering
	\includegraphics[width=0.95\linewidth]{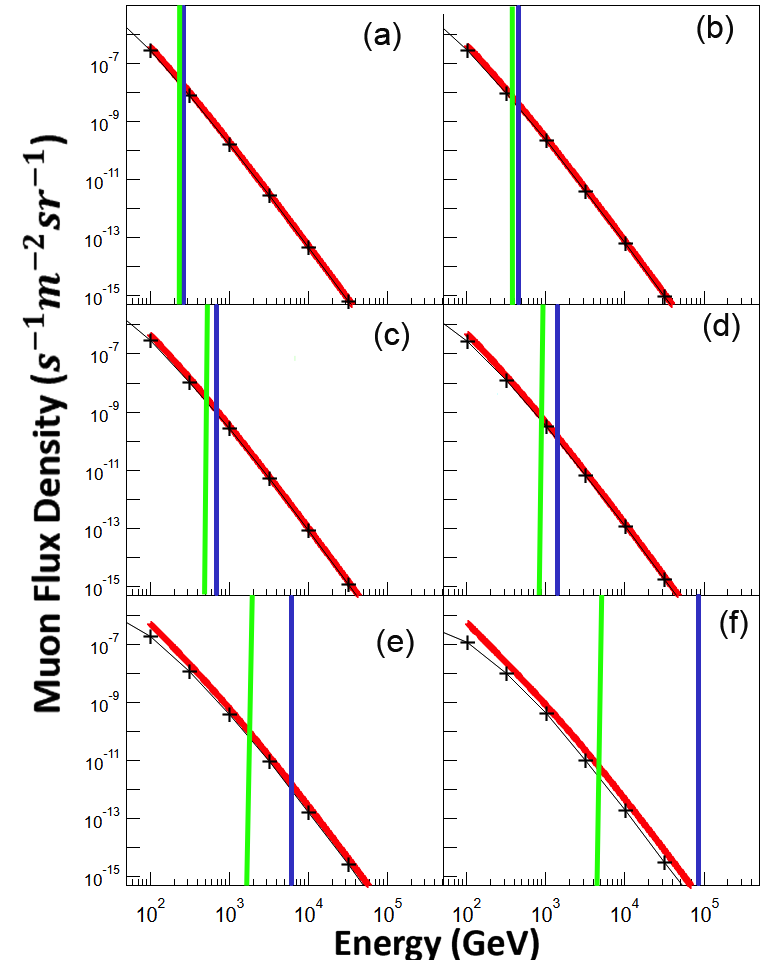}
	\caption{Sea level cosmic ray muon flux density at a fixed cos($\theta$) (a) 0.6, (b) 0.4, (c) 0.3, (d) 0.2, (e) 0.1, (f) 0.05, comparing Eq. \ref{equ:Surfcosmic} (red line) and  the measured data (black points). The blue vertical bars indicate the muon average energy cut-off or range considering both energy loss from ionization and radiation and at the experiments depth of 583 m.w.e.. The green vertical bars indicate the lower limit of muon energy cut-off only considering ionization energy loss at 583 m.w.e.
}
	\label{fig:compare}
\end{figure}

The slant angle calculated flux and experimental flux is in agreement in the high energy region below the muon cutoff energy or range, shown as a horizontal line (blue) in Figure \ref{fig:compare} and listed in Table \ref{ta:MuonECut}. The highest energy events estimated to be observed by the detector is $\sim$ 10 TeV, limiting the slant angle to cos($\theta)> $0.05. The curvature of the Earth was taken into account in these calculations.

Figure \ref{fig:compare} shows the cosmic ray muon spectrum predicted by the sea level muon density Formula \eqref{equ:Surfcosmic} and from measured data  \cite{lipari1991propagation} as a function of cos$\theta$.  cos$\theta$ is chosen as 0.6, 0.4, 0.3, 0.2, 0.1 and 0.05. Only the muons in the part to the right of the blue bar are able to pass through the rock and propagate to 583 m.w.e.. Thus from the Figure \ref{fig:compare}, the muon flux density predicted by Formula \eqref{equ:Surfcosmic} matches well with the experimental data. 

 \begin{table}
  \begin{center}
    \caption{Muon energy cutoff at 583 m.w.e as a function of muon slant angle.  The $E_{\mu}$ cutoff at each incident angle is the energy at which the muon exactly loses all the energy when passing through $583/ cos\theta$ m.w.e.. The curvture of the earth is considered in the calculation of the average cutoff energy (considering the energy loss from both ionization and radiation) and cutoff lower limit (only considering energy loss from ionization). \\} 
      \scalebox{0.9}
    { 
    \begin{tabular}{c c c c c}
    \hline
    $\mathbf{ cos ~ \theta}$ &~~ $\mathbf{\theta}$ & ~~$\mathbf{ 100 / cos \theta}$ &~~\bf{E$_\mu Average$ } &~~ \bf{E$_\mu Ionization$}\\
    &&(GeV)& \bf{Cutoff} (GeV) & \bf{Cutoff} (GeV)\\
    \hline 
    1.00  &~~ $0.0^\circ$  & 100  & 149 & 137 \\
    0.60  &~~ $53.1^\circ$ & 167  & 275 & 234\\
    0.40  &~~ $66.4^\circ$ & 250  & 463 & 358\\
    0.30  &~~ $72.5^\circ$ & 333  & 695 & 484\\
    0.20  &~~ $78.5^\circ$ & 500  & 1330 & 740\\
    0.10  & ~~$84.3^\circ$ & 1000 & 6045 & 1526\\
    0.05  &~~ $87.1^\circ$ & 2000 & 87800 & 3144\\
    0.00  &~~ $90.0^\circ$ &  -   & $> 10^6$ &  $> 10^6$\\
    \hline
    \end{tabular}
    }

  \label{ta:MuonECut}
  \end{center}
\end{table}

Although Formula \eqref{equ:Surfcosmic} is only valid when $E_\mu > 100/cos\theta\ GeV$ and $\theta < 70^\circ$, agian, muons not in this region will not pass through the thick rock layer to a depth of 583 m.w.e.. From this the underground cosmic ray spectrum can be found by propagating the sea level muon spectra using Formula \eqref{equ:Surfcosmic}.

\section {CUPP Muon Flux Density as a Function of depth} \label{app:CUPP_depth}
\begin{table}[h!]
\caption{Muon Flux Density as a function of depth measured in CUPP \cite{enqvist2005measurements}.}
\centering
\begin{tabular}{c c}
Depth        & Flux Density    \\
{[}m.w.e.{]} & {[}$m^{-2}s^{-1}${]}    \\ \hline
0            & $180 \pm 20 $      \\ 
210          & $1.3 \pm 0.2 $      \\ 
420          & $ (2.3 \pm 0.3) \times 10^{-1}$ \\ 
980          & $ (2.1 \pm 0.2) \times 10^{-2}$ \\ 
1900         & $ (3.2 \pm 0.3) \times 10^{-3}$ \\ 
2810         & $ (6.2 \pm 0.6) \times 10^{-4}$ \\ 
3960         & $ (1.1 \pm 0.1) \times 10^{-4}$ \\ \hline
\end{tabular}
\end{table}

\bibliography{Neutron.bib}

\end{document}